# Vortices, turbulence, and center of pressure in flow over pitching swept wings

Dipan Deb[+], Yuanhang Zhu[°], Philip Gaudio[+], and Kenneth Breuer[+]
[+] *Center for Fluid Mechanics, School of Engineering, Brown University, Providence, RI-02912, USA*
[°] *Department of Mechanical Engineering, University of California, Riverside, CA-92521, USA*

**This study examines the center of pressure (CoP) movement of rigid pitching swept wings based on prior measurements by Zhu and Breuer [1]. The wings analyzed feature sweep angles of $0°$, $10°$, and $20°$, and are subjected to large amplitude sinusoidal pitching instabilities below a critical torsional spring stiffness. The CoP location is determined from time-resolved force and moment measurements, revealing minimal variation in the cross-chord direction but significant spanwise and chord-wise movement, varied by sweep angle. The trajectory of the CoP varies with sweep angle due to the evolving strength and dynamics of the leading edge and tip vortices. The Force Moment Partitioning Method (FMPM) is applied to stereo Particle Image Velocimetry (PIV) data to identify contributions from wing kinematics, vortex structures, and viscous effects. This approach elucidates the roles of leading edge and tip vortices, as well as the periodic and stochastic components of the flow field, in influencing the net forces and moments.**

## I. Introduction

Swept wings undergo unsteady aerodynamic loading in various applications like the V-22 Osprey Aircraft [2] and in some Wind Turbines [3]. Moreover, swept wings are observed in flapping flight([4–6]) as well as in swimming ([7, 8]) in nature. Polhamus [9] showed that sweep contributes to spanwise vorticity transport, which enhances the lift in a delta wing by stabilizing the leading edge vortex (LEV). Chiereghin et al. [10] showed that there is significant spanwise flow (both towards the root and the tip of the wing) for a high aspect ratio plunging swept wing. However, spanwise flow due to sweep angle is not a ubiquitous strong criterion for LEV stabilization. Beem et al. [11] experimentally demonstrated that despite having a strong spanwise flow on a swept wing under plunging motion, the LEV breaks down and convects downstream. Wong et al. [12] underlines this claim by stating that two-dimensional spanwise flow without any velocity gradient is not sufficient to stabilize LEVs. The spanwise flow or vorticity has to have a gradient for stretching and convection.Wong and Rival [13] demonstrated experimentally that LEV stability can be improved by draining vorticity through spanwise vorticity transport, which is assisted by profile sweep. Furthermore, they supported the claim of Chiereghin et al. [10] regarding the positive role of high aspect ratio in LEV stability. Onoue and Breuer [14] proposed a universal scaling for the LEV formation time as a function of the reduced pitching frequency, the sweep angle, and the pivot location with respect to the chord. Moreover, they claimed that the LEV stability can be improved by vorticity annihilation, which is primarily assisted by the leading edge sweep, strengthening the assertions of Wojcik and Buchholz [15] and Wong and Rival [13]. Visbal and Garmann [16] numerically assessed the effect of sweep angles on the dynamic stall of pitching airfoils and claimed that the leading edge arch vortex structures and their stability are affected by sweep angle. A similar arch vortex structure and the effect of sweep angle are also observed by Zhu and Breuer [1] in their elastically mounted pitching wing experiment. Moreover, both studies ([1, 16]) reported the independence of lift forces from sweep angle, but its dependence on the pitching moment. This observation is one of the motivations for the present study.

Besides the leading edge vortex (LEV), the Wing Tip Vortex (WTV) plays a significant role in the aerodynamic characteristics of a finite span three-dimensional wing. The studies, such as Taira and Colonius [17], Kim and Gharib [18], Hartloper et al. [19] suggest that the onset of LEV and WTV is independent of each other. However, the studies by Hartloper et al. [19], Birch and Dickinson [20] also claim that WTV can potentially hold the LEV and delay its shedding. Studies like Taira and Colonius [17], Zhang et al. [21, 22], Ribeiro et al. [23] showed that the WTV affects the unsteady wake of both unswept and swept wings. In the present study, we have also noticed a huge contribution of WTV in the normal force and rolling moment.

In any aerodynamic or hydrodynamic application, flight stability is an important consideration, as well as how flow structures like LEV or WTV affect this stability. The center of pressure (CoP) is an important parameter to determine such stability. In a steady flow, CoP can be experimentally explicitly obtained using a center of pressure apparatus [24]. However, in an unsteady flow, the center of pressure moves, and the determination of its movement requires accurate measurement of the unsteady forces and moments acting on the wing. Connecting those forces with flow structures requires simultaneous measurement of the surrounding flow, and such coupled measurements can be extremely challenging.

An appealing simplification is to compute the aerodynamic forces from the surrounding flow field. Chorin [25] is the first work that laid the foundation of obtaining the pressure field from velocity data by reducing the Navier-Stokes equation to a Poisson equation of pressure. Adrian et al. [26] talked about obtaining the spatial velocity gradient of experimentally obtained velocity data, which is crucial for accurately calculating pressure. De Kat and Van Oudheusden [27] proposed an algorithm to calculate the pressure field from the quasi-steady or unsteady velocity field, and they applied this technique in turbulent flows. Charonko et al. [28] proposed a velocity error propagation mitigation technique in the pressure estimation using proper orthogonal decomposition (POD). Dabiri et al. [29] utilized median polling of several integration paths through the pressure gradient field to reduce the effect of measurement errors of the velocity field in the determination of the pressure field. Recently Chen et al. [30] applied a data-driven approach to obtain the pressure field using snapshot PIV and fast probes. However, these methods fail to capture the contributions of different flow structures, added mass, viscous effects, etc.

A very promising complement to these methods for computing the pressure field is the Force and Moment Partitioning Method (FMPM), which was first proposed by Quartapelle and Napolitano [31] and later utilized by several researchers [1, 32–38]. Due to its flexibility in handling a wide range of applications, like the vortex-dominated flows, this technique has seen increased attention in recent years.

In this method, an "influence potential", $\phi$, is defined that satisfies the Laplace equation with certain Neumann boundary conditions that depend on the direction of the force or moment being estimated. Thereafter, the Navier-Stokes equation is projected on the gradient of the influence potential, $\nabla \phi$, which enables us to quantify the contributions from the vortical flow structures, added mass, viscosity, and the outer boundary term. Menon and Mittal [36], Menon et al. [39], Zhang and Taira [40] utilized this method on two-dimensional and three-dimensional flow to investigate the force distributions resulting from vorticity on swept wings.

This method has been applied to experimental PIV data as well by Zhu et al. [41] to analyze the effect of vortices shed by a 2D pitching airfoil in quiescent fluid on the pitching moment. It should be mentioned that the vortical term is the only non-linear term in the FMPM method. Zhu et al. [41] showed that in a phase-averaged flow field, the vortical term may include a strong contribution from the fluctuating instantaneous velocity field. The present study investigates the contribution of the fluctuating term to the movement of the center of pressure.

In the present study, we have three goals: (i) to use FMPM to estimate the forces, moments, and the center of pressure acting on a pitching swept wing, (ii) to assess the role of turbulence and measurement noise on the accuracy of that estimation, and (iii) to connect those forces, moments and CoP to the coherent flow structures like LEV and WTV. To address these goals, we apply the FMPM method to previously-acquired stereoscopic PIV data [1] to obtain the force and moment distributions in the directions of interest. We also extend the methodology developed by Zhu et al. [41] to quantify the influence of the fluctuating velocity field on the CoP movement.

The paper is organized as follows: In the next section, we review the experimental data set of Zhu and Breuer [1]. The following section II discusses the experimental setup. Section III discusses the first goal, that is, the use of FMPM for force, moment, and CoP estimation on a pitching swept wing. Section IV aims to connect the forces, moments, and CoP to the coherent flow structures like LEV and WTV. Finally, section V summarizes the key findings of this study.

## II. Experimental Setup

Complete details of the experimental setup are given by Zhu and Breuer [1] but are summarized here for convenience. All the experiments were performed in a free-surface water tunnel, with a test section $W \times D \times L = 0.8m \times 0.6m \times 4.0m$, at Brown University. Figure 1(a) shows a schematic of the experimental setup. The cross-section of the wing is NACA0012. One endplate is attached to the root of the wing; however, the tip of the wing is free. The data is acquired with a six-axis force transducer (FT) to measure 3 forces and 3 moments, by Zhu and Breuer [1] as reported. An optical encoder is also utilized to measure the pitching angle of the wing. Three sweep angles (denoted by $\Lambda$, as shown in Figure 1(b)) are considered for this study, $\Lambda = 0°$, $\Lambda = 10°$ & $\Lambda = 20°$. All the wings have a chord length, $c = 0.1m$ & span $s = 0.3m$, which makes the aspect ratio of all the wings $AR = 3$. In all the cases mentioned in this study, the free stream velocity is kept consistent at $U = 0.3m/s$. Hence, the chord-based free stream Reynolds number is also consistent at $Re = \rho Uc/\mu = 30,000$, where $\rho$ is the density of water. The pitching motion is characterized as, $\theta = Asin(2\pi ft)$, where $A = 60°$ and $f_p = 0.255Hz$ are pitching amplitude and frequency respectively (reduced frequency, $\kappa = \pi fc/U = 0.27$). The mean pitching amplitude is also zero. The time period of the pitching oscillation is denoted by $T$ $(= 1/f_p)$, and in the current manuscript, the time parameter is normalized $(t/T)$ by the time period. The starting point $(t/T = 0)$ of the pitching motion is when $\theta = 0°$. The parameters of the pitching motion are consistent across all the sweep angles. The only difference between all the cases is the sweep angle. As shown in the Figure 1, the

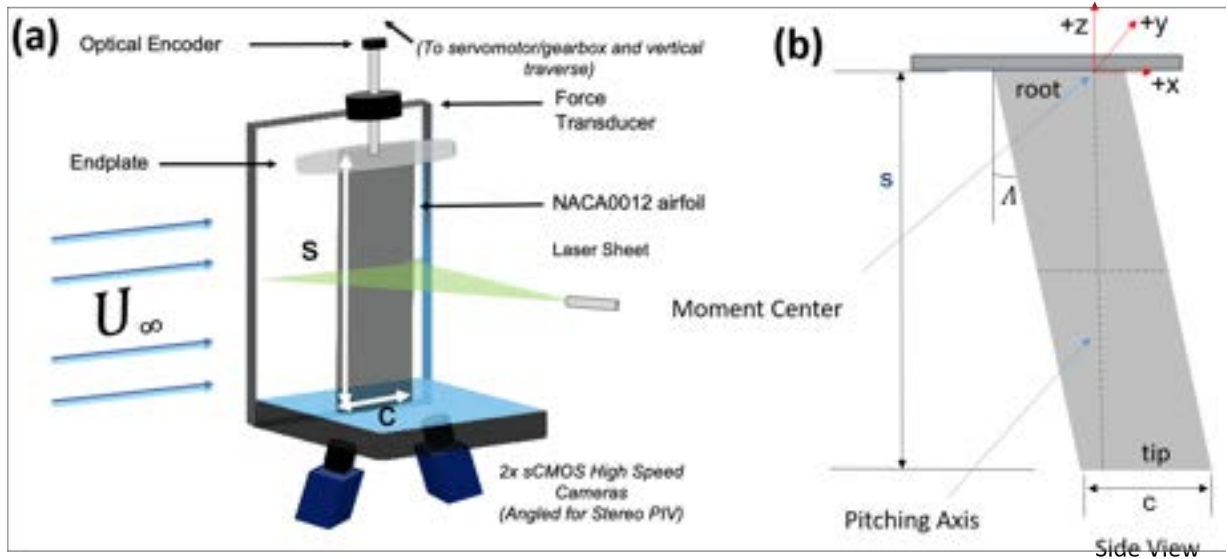

**Fig. 1** (a) A schematic of the experimental set-up. (b) A sketch of the side view of the wing. The pitching axis passes through the mid-chord and mid-span of the wings. Where the pitching axis meets the end plate is considered the moment center for the present study.

flow field is measured using stereo particle image velocimetry (PIV) using two cameras. The experiment provided us with 3 velocities in a plane. Since this is a stereo, not volumetric, PIV, the spanwise variation of the instantaneous velocity is not captured through this measurement. The wing was traversed vertically to obtain the flow field at different spanwise planes. For each PIV measurement, 25 pitching cycles are captured. Data from these 25 pitching cycles are phase-averaged for each plane, and data from each plane are considered to reconstruct the 3D flow field. This phase-averaged reconstructed 3D flow field has spanwise variation and is useful to calculate the velocity gradients in all directions discussed later in the present work. However, the spanwise velocity gradient of the instantaneous velocity field could not be calculated due to the planar PIV measurement. Figure 1(b) shows the pitching axis, which passes through the mid-chord and mid-span of the wing. The Figure 1(b) also shows the moment center, about which all the moments are calculated in this study. This point is also the origin of our coordinate system.

## III. Methods: CoP Calculation and FMPM

**A. Force and Moment Measurement**

The time-varying forces and moments were measured by Zhu and Breuer [1] using a 6-axis Force Torque (FT) transducer. The FT measures forces and moments in the $x$, $y$ and $z$ directions, which are the chord-wise, cross-chord-wise, and span-wise directions, respectively, in the coordinate system relative to the wing (Fig. 2(a)). The drag, $D$, is in the direction of the flow, and the lift, $L$, is perpendicular to that direction. Hence, to obtain lift and drag, the measured forces $F_y$ & $F_x$ are transformed according to the pitching angle $\theta$:

$$L = F_y Cos\theta - F_x Sin\theta \tag{1}$$

$$D = F_y Sin\theta + F_x Cos\theta. \tag{2}$$

The lift and drag forces are normalized to calculate the force coefficients:

$$C_L = \frac{L}{\frac{1}{2}\rho U^2 sc} \qquad C_D = \frac{D}{\frac{1}{2}\rho U^2 sc}. \tag{3}$$

Measurement in the $z$ or binomial direction is not affected by the pitching motion. Hence the acquired pitching moment $M_z$ is normalized as follows:

$$C_{M_z} = \frac{M_z}{\frac{1}{2}\rho U^2 sc^2}. \tag{4}$$

The phase-averaged time-varying Lift and Drag coefficients are shown in Figure 2(b) for pitching amplitude $A = 60°$ and frequency $f_p = 0.255 Hz$ (reduced frequency, $\kappa = 0.27$), which is also reported by Zhu and Breuer [1]. Solid lines represent the lift force acting on the wing, while dashed lines indicate drag. The aerodynamic forces corresponding to sweep angles of $\Lambda = 0°$, $\Lambda = 10°$, and $\Lambda = 20°$ are shown in blue, red, and green, respectively, a color scheme consistently applied throughout this manuscript for the sweep angles. Notably, the lift and drag coefficients exhibit minimal dependency on the sweep angle. The pitching moment coefficients for various sweep angles are presented using the same color scheme as the force measurements in Figure 2(c). Unlike the force coefficients, the pitching moment does demonstrate a dependency on the sweep angle, which motivated the calculation of the center of pressure. Additionally, due to the symmetry of the pitching motion, the force and moment measurements exhibit symmetry about the midpoint of the oscillation cycle. Hence, all subsequent analyses will focus exclusively on the half-cycle.

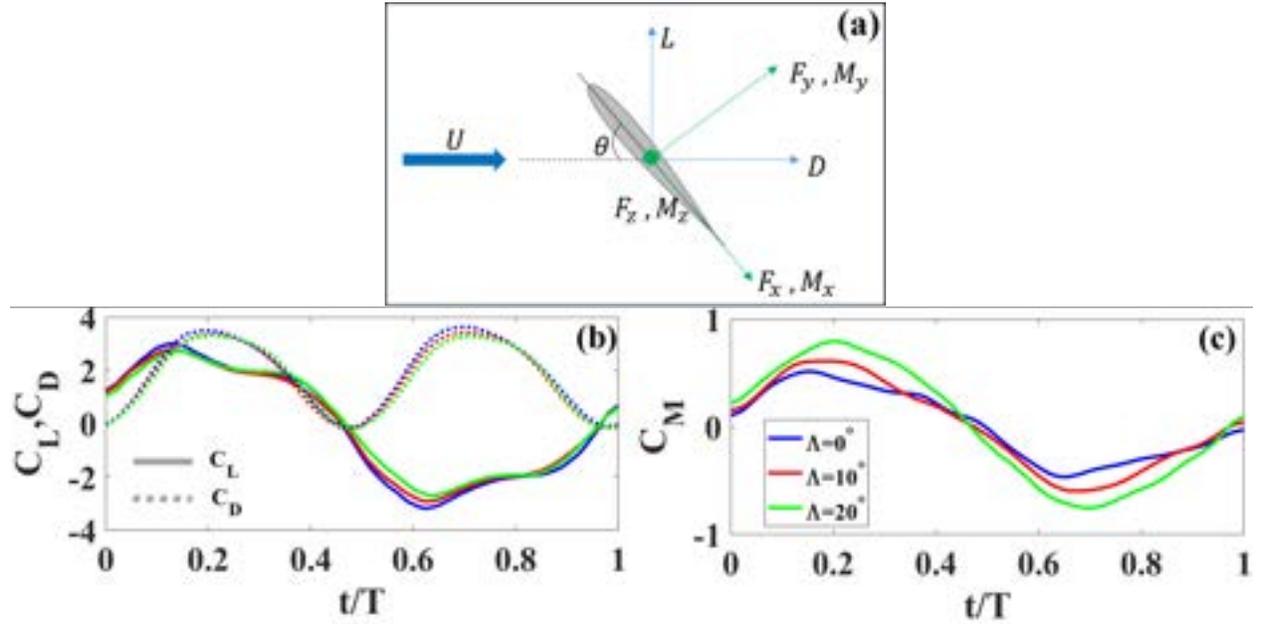

Fig. 2  (a) Schematic of the force and moment directions with pitching angle (b) Measured phase-averaged Lift and Drag coefficients. Solid line denotes lift, and the dashed line represents drag. (c) Measured pitching moment coefficient for various sweep angle $\Lambda$

## B. Center of Pressure Calculation

The center of pressure (CoP) is calculated using the basic definition of moment, $\vec{r} \times \vec{F} = \vec{M}$. In the present case, $\vec{F}$ & $\vec{M}$ are measured from the 6-axis FT. Hence, the calculation of $\vec{r}$ can reveal the location of CoP. Since there are 3 forces and 3 moments, this equation can be written in a matrix form as follows,

$$\begin{bmatrix} 0 & F_z & -F_y \\ -F_z & 0 & F_x \\ F_y & -F_x & 0 \end{bmatrix} \begin{pmatrix} x_{cp} \\ y_{cp} \\ z_{cp} \end{pmatrix} = \begin{Bmatrix} M_x \\ M_y \\ M_z \end{Bmatrix}. \tag{5}$$

The matrix equation 5 is of the form $[F](r) = (M)$, where $[F]$ & $(M)$ are known from the measurement and $\vec{r} = x_{cp}\vec{i} + y_{cp}\vec{j} + z_{cp}\vec{k}$ needs to be evaluated. Here, $\vec{i}, \vec{j}$ & $\vec{k}$ are unit vectors in the direction of x, y & z respectively.

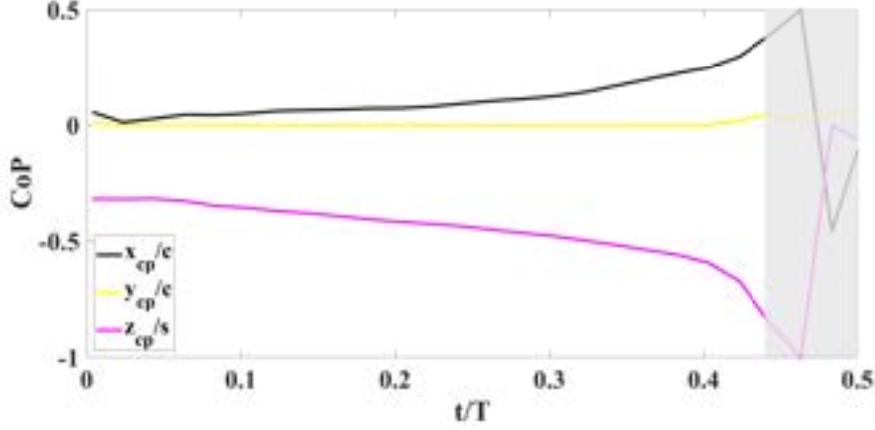

**Fig. 3** Normalized movement of the CoP in all three directions for sweep angle $\Lambda = 0°$

However, the $[F]$ matrix shown in equation 5 is a $3 \times 3$ skew-symmetric matrix, which by definition is a singular matrix. Thus, this equation can not be solved by performing simple matrix inversion. We utilize the optimization technique by minimizing the error, $f(r) = |[F](r) - (M)|$, at every time step to obtain the location of CoP. An example solution for sweep angle $\Lambda = 0°$ is shown in Figure 3. which illustrates the movement of the center of pressure (CoP) in the chord-wise ($x$), cross-chord-wise ($y$), and span-wise ($z$) directions, represented by black, yellow, and purple curves, respectively. The movement is shown for half a cycle, leveraging symmetry. The CoP displacements in the $x$ and $y$ directions are normalized by the chord length, while the span-wise movement is non-dimensionalized by the span length. It is evident that the CoP movement in the cross-chord-wise ($y$) direction is negligible compared to its movement in the $x$ and $z$ directions. This behavior can be attributed to the symmetric geometry of the NACA0012 airfoil section. This observation is consistent across all cases studied in this study. Hence, this manuscript focuses exclusively on CoP movement in the $x$ and $z$ directions. By setting $y_{cp} = 0$ in the matrix formulation 5 we obtain,

$$-F_y z_{cp} = M_x \tag{6}$$

$$F_y x_{cp} = M_z. \tag{7}$$

From equations 6 & 7, it can be seen that the dominant forces and moments influencing CoP movement are the normal force ($F_y$), rolling moment ($M_x$) and pitching moment ($M_z$). At $t/T > 0.45$ the forces and moments approach zero, which reduces the confidence in CoP calculation in this regime. To emphasize this uncertainty at these times, in Figure 3 (and subsequent figures) this region has been shaded.

## C. Use of Force/Moment Partitioning

To explain the movement of CoP, it is necessary to know the variation of the primary contributing forces (normal) and moments (pitching and rolling) in the corresponding directions. We utilize the Force Moment Partitioning Method (FMPM) [31] to acquire the required variation in force and moments. We briefly review the method, but more details can be found in [1, 42].

The application of FMPM requires calculation of an "influence potential", $\phi$, that satisfies the Laplace equation with two Neumann boundary conditions on the wing body and the outer boundary as follows:

$$\nabla^2 \phi_i = 0, \qquad \frac{\partial \phi}{\partial n} = \begin{cases} \left[ (\mathbf{x} - \mathbf{x}_p) \times \mathbf{n} \right] \cdot \mathbf{e}_i, & \text{on the aerofoil for moment,} \\ 0, & \text{on the outer boundary,} \end{cases} \qquad (8)$$

$$\nabla^2 \phi_i = 0, \qquad \frac{\partial \phi}{\partial n} = \begin{cases} \mathbf{n}_i, & \text{on the aerofoil for force,} \\ 0, & \text{on the outer boundary,} \end{cases} \qquad (9)$$

Here $i$ denotes the direction of the force or moment, and $\mathbf{n}$ denotes the unit vector normal to the surface. The equation 9 shows the boundary condition of the influence potential for forces, and equation 8 denotes the boundary condition for moments. $\mathbf{x} - \mathbf{x}_p$ shows the location vector pointing from the moment center. Figure 4 shows an example of the influence potential, calculated for rolling moment ($M_x$) with sweep angle $\Lambda = 0°$.

In the present case, we are only interested in the normal force, pitching moment, and rolling moment. Influence Potentials for all the forces and moments across all the sweep angles, discussed in the present study, are shown in Figure 14 in the appendix. With respect to the wing, the directions of the force and moments are unchanged during the pitching cycle. Hence, the influence potential needs to be calculated only once. This influence potential identifies the spatial contribution of flow structures on the resultant force/moment. With the help of the influence potential and the flow field data, FMPM can help us identify contributions arising from added mass ($C_{F_i}^\kappa / C_{M_k}^\kappa$), vortical flow structures $(C_{F_i}^\omega/C_{M_k}^\omega)$, viscous $(C_{F_i}^\sigma/C_{M_k}^\sigma)$, and boundary terms $(C_{F_i}^\Sigma/C_{M_k}^\Sigma)$. For forces, these terms are given by:

$$C_{F_i} = C_{F_i}^\omega + C_{F_i}^\kappa + C_{F_i}^\sigma + C_{F_i}^\Sigma \tag{10}$$

$$C_{F_i}^\omega = \int_V -2Q\phi_i dV \tag{11}$$

$$C_{F_i}^\kappa = \int_B -\hat{n}.(\frac{d\vec{U_b}}{dt}\phi_i)dS \tag{12}$$

$$C_{F_i}^\sigma = \int_B \frac{1}{Re}\{(\vec{\omega}\times\hat{n}).\nabla\phi_i - (\vec{\omega}\times\hat{n}).\hat{e}_i\}dS \tag{13}$$

$$C_{F_i}^\Sigma = \int_\Sigma [-\hat{n}.(\frac{d\vec{u}}{dt}\phi_i) + \frac{1}{Re}(\vec{\omega}\times\hat{n}).\nabla\phi_i]dS. \tag{14}$$

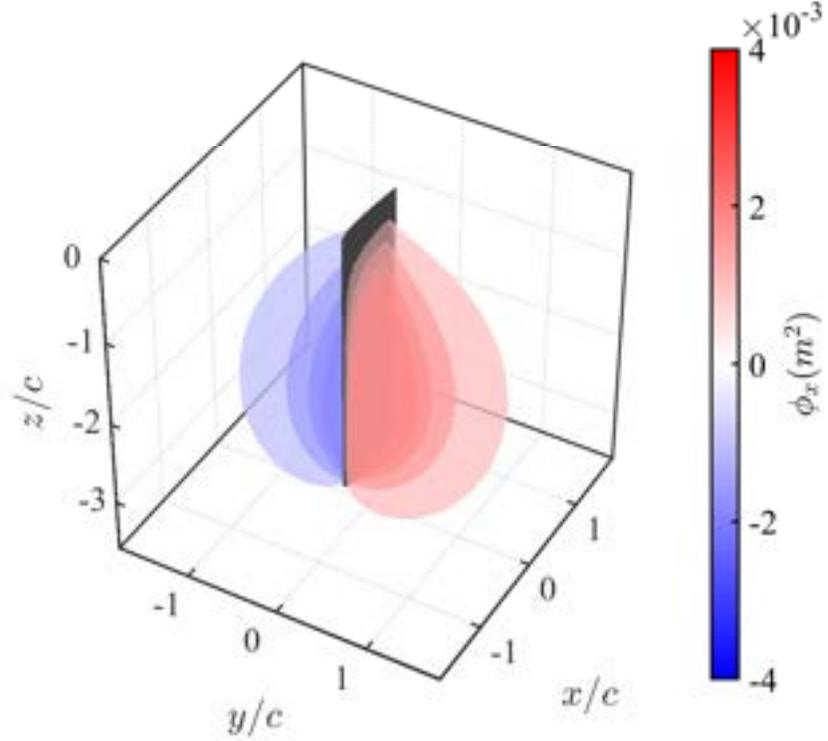

**Fig. 4 Influence Potential of Rolling Moment coefficient for sweep angle $\Lambda = 0°$**

For moments, the corresponding terms are:

$$C_{M_k} = C_{M_k}^\omega + C_{M_k}^\kappa + C_{M_k}^\sigma + C_{M_k}^\Sigma \tag{15}$$

$$C_{M_k}^\omega = \int_V -2Q\phi_k dV \tag{16}$$

$$C_{M_k}^\kappa = \int_B -\hat{n}.(\frac{d\vec{U_b}}{dt}\phi_k)dS \tag{17}$$

$$C_{M_k}^\sigma = \int_B \frac{1}{Re}\{(\vec{\omega}\times\hat{n}).\nabla\phi_k - (\vec{\omega}\times\hat{n}).[\hat{e}_k\times(\vec{x}-\vec{x}_p)]\}dS \tag{18}$$

$$C_{M_k}^\Sigma = \int_\Sigma [-\hat{n}.(\frac{d\vec{u}}{dt}\phi_k) + \frac{1}{Re}(\vec{\omega}\times\hat{n}).\nabla\phi_k]dS \tag{19}$$

In the above equations, *i* & *k* represent the direction of forces and moments, respectively. The vortical terms depend on $Q$, which is the 2nd invariant of the velocity gradient tensor:

$$Q = \frac{1}{2}(||\mathbf{\Omega}||^2 - ||\mathbf{S}||^2), \tag{20}$$

where $\mathbf{\Omega}$ is the rotational tensor and $\mathbf{S}$ is the strain-rate tensor.

### D. Role of Turbulence and Measurement Noise

Calculating the vortical term requires the calculation of $Q$ from the PIV measurement of the velocity field. This calculation utilizes the velocity gradients of the measured velocity field. Now, the velocity field can be resolved into the phase averaged and the fluctuating component and so is $Q$. Zhu et al. [41] resolved the vortical term into time average, phase-averaged, and fluctuating terms. However, in the present work, we resolved it into only phase-averaged and fluctuating terms as follows,

$$\vec{u}(t) = \langle\vec{u}\rangle(t/T) + \vec{u}'(t) \tag{21}$$

Where $\langle u \rangle$ denotes the phase-averaged and $u'$ denotes the fluctuating component of the velocity field. Now, Q takes the following form after resolving equation 20,

$$Q = -\frac{\partial u}{\partial y}\frac{\partial v}{\partial x} - \frac{\partial v}{\partial z}\frac{\partial w}{\partial y} - \frac{\partial u}{\partial z}\frac{\partial w}{\partial x} + \frac{\partial u}{\partial x}\frac{\partial v}{\partial y} + \frac{\partial v}{\partial y}\frac{\partial w}{\partial z} + \frac{\partial u}{\partial x}\frac{\partial w}{\partial z} \tag{22}$$

Plugging in the phase-averaged and fluctuating component of the velocity into the formulation of Q, we can resolve it in the following way,

$$\langle Q \rangle = Q_{\langle\langle u\rangle,\langle u\rangle\rangle} + Q_{\langle u',u'\rangle} \tag{23}$$

Where,

$$Q_{\langle\langle u\rangle,\langle u\rangle\rangle} = \langle -\frac{\partial \langle u\rangle}{\partial y}\frac{\partial \langle v\rangle}{\partial x} - \frac{\partial \langle v\rangle}{\partial z}\frac{\partial \langle w\rangle}{\partial y} - \frac{\partial \langle u\rangle}{\partial z}\frac{\partial \langle w\rangle}{\partial x} + \frac{\partial \langle u\rangle}{\partial x}\frac{\partial \langle v\rangle}{\partial y} + \frac{\partial \langle v\rangle}{\partial y}\frac{\partial \langle w\rangle}{\partial z} + \frac{\partial \langle u\rangle}{\partial x}\frac{\partial \langle w\rangle}{\partial z} \rangle \tag{24}$$

$$\text{and, } Q_{\langle u',u'\rangle} = \langle -\frac{\partial u'}{\partial y}\frac{\partial v'}{\partial x} - \frac{\partial v'}{\partial z}\frac{\partial w'}{\partial y} - \frac{\partial u'}{\partial z}\frac{\partial w'}{\partial x} + \frac{\partial u'}{\partial x}\frac{\partial v'}{\partial y} + \frac{\partial v'}{\partial y}\frac{\partial w'}{\partial z} + \frac{\partial u'}{\partial x}\frac{\partial w'}{\partial z} \rangle \tag{25}$$

$$\tag{26}$$

Now, $Q_{\langle\langle u\rangle,\langle u\rangle\rangle}$ is calculated using the phase-averaged velocity field, and $Q_{\langle u',u'\rangle}$ is from the fluctuating velocity field. Since the velocity data is obtained using Stereo PIV (measuring $u, v, w$), scanning horizontal planes, we are only able to evaluate $z$-derivatives of the phase-averaged velocities, but not of the instantaneous velocity fields. Thus, although the full form of $Q_{\langle\langle u\rangle,\langle u\rangle\rangle}$ can be calculated, the quantities $\frac{\partial u'}{\partial z}$, $\frac{\partial v'}{\partial z}$ & $\frac{\partial w'}{\partial z}$ could not be calculated using the current data. In the present analysis these terms are assumed to be small, and the justification for this is provided in Appendix B The present study is limited to $Q^{2D}_{\langle u',u'\rangle} = \langle -\frac{\partial u'}{\partial y}\frac{\partial v'}{\partial x} + \frac{\partial u'}{\partial x}\frac{\partial v'}{\partial y}\rangle$ only. The present study is concluded utilizing $Q^{3D}_{\langle\langle u\rangle,\langle u\rangle\rangle}$ and $Q^{2D}_{\langle u',u'\rangle}$. When these two components are applied to the vortical term, we get,

$$C^{\omega}_M \text{ (total)} = -2\int_V Q\phi_i\, dV \tag{27}$$

$$= -2\int_V (Q^{3D}_{\langle\langle u\rangle,\langle u\rangle\rangle} + Q^{2D}_{\langle u',u'\rangle})\phi_i\, dV \tag{28}$$

$$C^{\omega}_M \text{ (total)} = \langle C^{\omega}_M\rangle \text{ (average)} + C^{\omega'}_M \text{ (fluctuating)} \tag{29}$$

For the remainder of this manuscript, the total vortical term is partitioned into two parts: the phase-averaged and the fluctuating component.

## IV. Results and Discussion

### A. Comparison of different FMPM terms

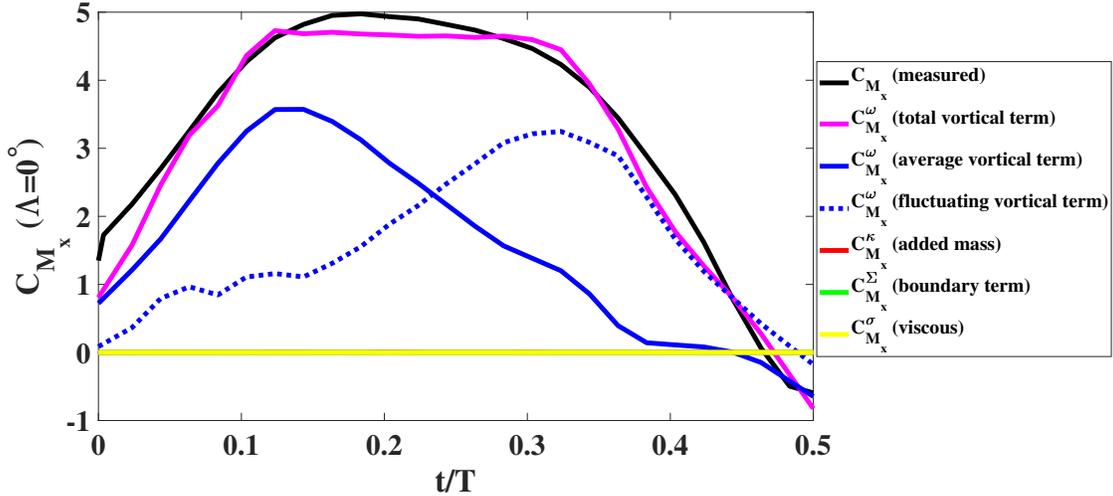

**Fig. 5** Comparison of all the FMPM terms with measured Rolling Moment coefficient for sweep angle $\Lambda = 0°$

Figure 5 shows an example of the FMPM estimation of the rolling moment coefficients for different terms compared to the measured value. It also shows a breakdown of the total vortical term into average and fluctuating terms of the rolling moment coefficient ($C_{M_x}$) for sweep angle $\Lambda = 0°$. The solid black line represents the measured rolling moment coefficient by FT. The solid purple and blue curves denote the total and averaged vortical term. Although $Q^{3D}_{\langle\langle u\rangle,\langle u\rangle\rangle}$ and $Q^{2D}_{\langle u',u'\rangle}$ are utilized to calculate the phase-averaged and fluctuating components of the vortical term, the FMPM calculation of the total vortical term reveals good agreement with the FT measurement. Furthermore, Figure 5 shows that, except for the vortical terms, all the other terms (added mass, viscous, and boundary terms) are negligible. The other terms are so small that they almost overlap with each other. This observation is consistent across all the forces, moments, and sweep angles. Hence, the current study only discusses the vortical terms.

One can ask, how is the added mass term negligible in an unsteady flow? Zhu et al. [41] showed that the added mass term is significant compared to the vortical term for a pitching airfoil in quiescent water. The added mass term in the FMPM method is not dependent on the free stream velocity; it only depends on the kinematics of the wing [35]. Hence, with similar kinematics, cases with or without free stream flow would have similar added mass terms. However, with a free stream, the vortical terms become more significant in vortex-dominated flows. Zhu and Breuer [1] showed the dominance of the vortical terms in such cases, where the kinematics of the pitching wing was similar to their 2023 work [41]. This has also been observed in the present study, where the free stream induces strong leading-edge and wingtip vortices, rendering the added mass term negligible in comparison.

## B. Comparison between the Measured Force/Moments and FMPM

The subsection III.B mentions that the primary forces and moments that contribute to the movement of CoP are Normal Force ($F_y$), Rolling Moment ($M_x$) & Pitching Moment ($M_z$). Hence, in order to understand the CoP dynamics, we applied FMPM on these forces and moments. As stated in the previous subsection, only the vortical terms are significant.

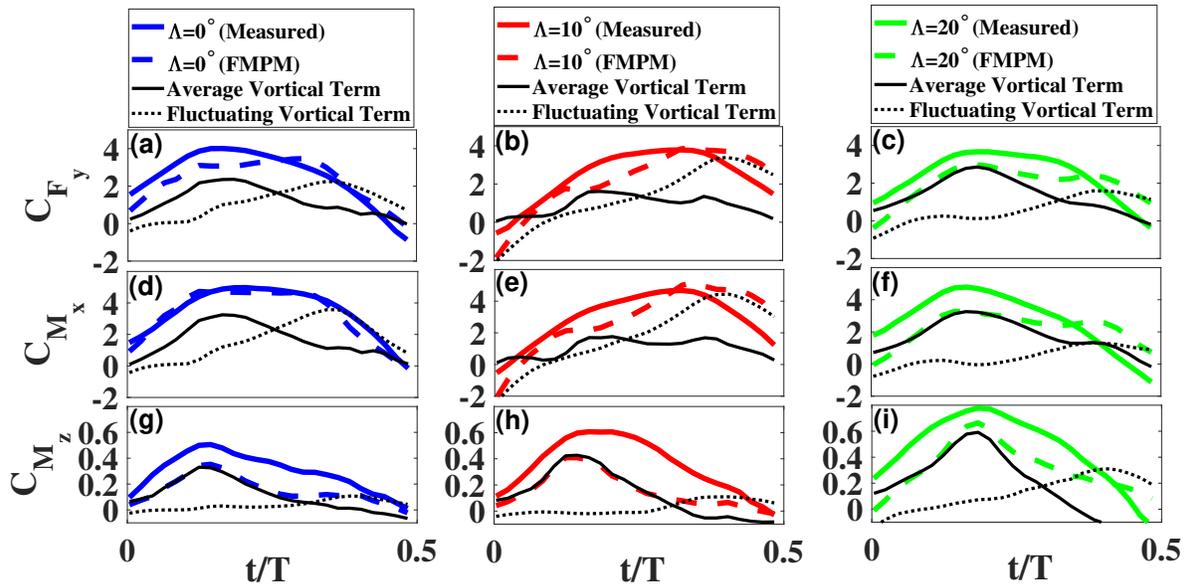

Fig. 6  Comparison between the coefficients obtained from FMPM and the measured values of (a-c) Normal Force ($C_{F_y}$), (d-f) Rolling Moment ($C_{M_x}$) & (g-i) Pitching Moment ($C_{M_z}$) for different sweep angles. The solid and dotted black lines represent the averaged and fluctuating vortical terms for the corresponding force and moment.

Figure 6 shows the comparison between the total vortical term estimation with the relevant forces and moments with their measured values. The 1st, 2nd, and 3rd row show the Normal Force ($C_{F_y}$), Rolling Moment ($C_{M_x}$), and Pitching Moment coefficients, ($C_{M_z}$) respectively. These forces and moments are normalized by the same quantity shown in equations 3 & 4. Similarly, the 1st, 2nd, & 3rd column or the blue, red & green colors denotes the sweep angle $\Lambda = 0°$, $\Lambda = 10°$, & $\Lambda = 20°$ respectively. The plots show only the half-cycle due to symmetry. The solid lines denote the measured values of the forces/moments, and the dashed lines show the FMPM estimation. Moreover, in each plot, solid and dotted black lines show the averaged and fluctuating vortical terms for the corresponding plot. Although not perfect, the plots reveal a good match between the measured forces/moments and the FMPM estimation. It should be highlighted that both average and fluctuating terms have significant contributions to the total vortical term. Depending on the time instances during the pitching cycle, these contributions direct the CoP movement, which is discussed in section IV.

## C. CoP Movement Estimation with FMPM

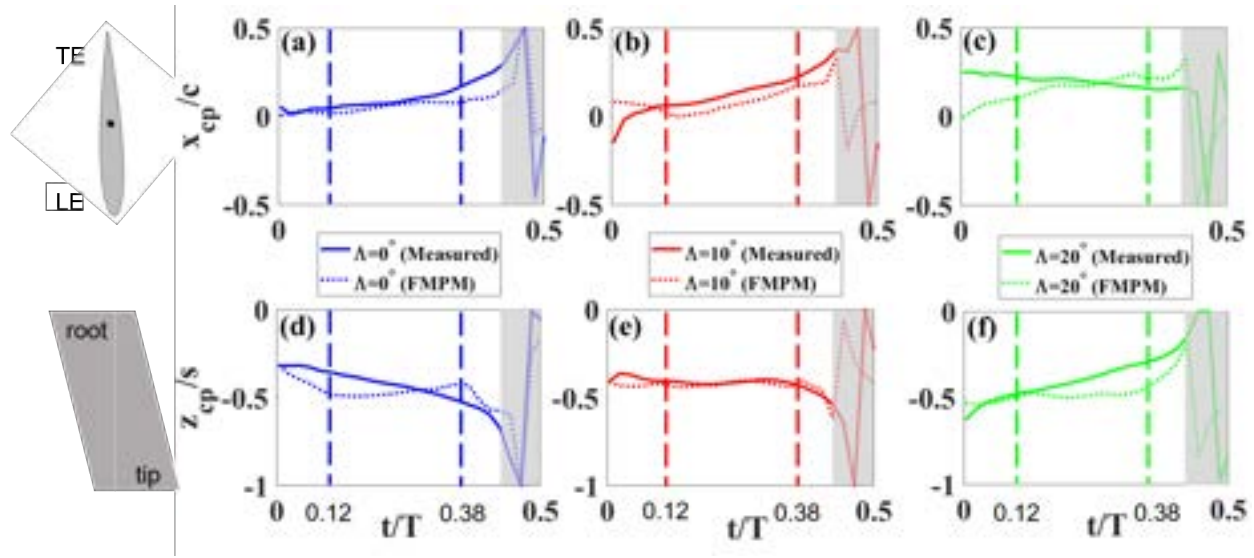

**Fig. 7** The figure shows the comparison between the movement of CoP calculated using the FT measured forces & moments and total vortical terms from the FMPM method in the (a) chord-wise or x-direction (b) span-wise or z-direction

We have already seen (Sec. III.B) that the CoP movement in the $y$-direction is negligible. Hence, the CoP movement in the $x$ and $z$ directions is compared, which were calculated from the measured forces/moments and the FMPM method, is shown in Figure 7. The vortical terms shown in Figure 6 have been plugged into equation 6 & 7 to obtain the locus of CoP. 1st row depicts the CoP movement in the $x$-direction. For better comprehension, a cross-section of the NACA0012 airfoil is shown on the left side of the figures. The color scheme is consistent with the previous figure, that is blue, red, and green represent the sweep angle $\Lambda = 0°$, $\Lambda = 10°$, & $\Lambda = 20°$, respectively. It can be noted that for sweep angles $\Lambda = 0°$ & $\Lambda = 10°$ the CoP moves towards the trailing edge (TE), whereas for $\Lambda = 20°$ the CoP is almost invariant.

Figure 7(d-f) shows the CoP movement in the $z$-direction. Similarly to the 1st row, a span-wise view of the wing is shown on the left side of the 2nd row, with the same color scheme for sweep angles. It can be observed that for $\Lambda = 0°$ & $\Lambda = 10°$ the CoP moves towards the tip, while it shifts towards the root of the wing for $\Lambda = 20°$. Again, while the agreement is not perfect, FMPM is able to show a similar trend for the locus of the CoP as calculated from the FT measurement, and reinforces the idea that FMPM can be a valuable tool for understanding the factors influencing CoP movement.

Furthermore, we chose two time instants: $t/T = 0.12$ & $t/T = 0.38$, to explore the connections between the force and moment coefficients and the corresponding flow structures. These two time instances are shown in every subplots in Figure 7 with dashed lines. They are chosen because both of them have almost the same pitching angle ($\theta \sim 40°$). However, the fluctuating vortical term is much more significant at $t/T = 0.38$ than it was at $t/T = 0.12$ (Figure 6 & 5), reinforcing the significance of the fluctuating vortical term in the CoP movement. In the following sections, we shall look into the force and moment distributions along $x$ & $z$ directions, to explain the CoP movement from $t/T = 0.12$ to $t/T = 0.38$.

### D. Connection between Flow Structures and CoP at $t/T = 0.12$

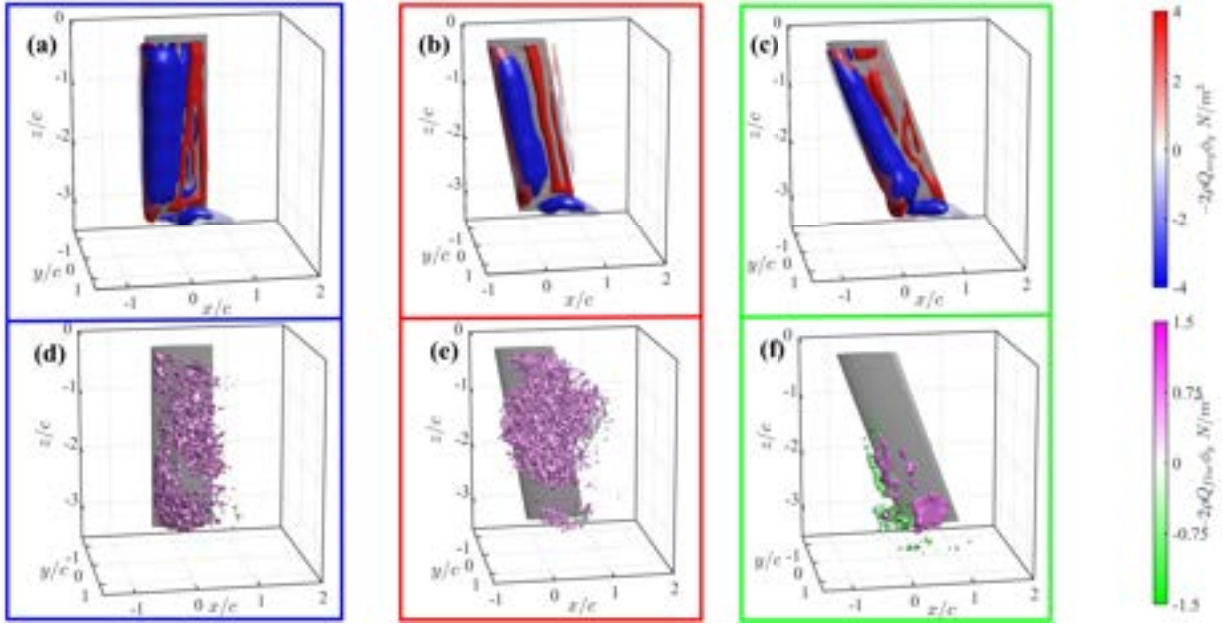

**Fig. 8 Iso surface plot of the average normal force density $-2\rho Q_{avg}\phi_y$ sweep angle (a) $\Lambda = 0°$ (b) $\Lambda = 10°$ (c) $\Lambda = 20°$ and the fluctuating normal force density $-2\rho Q_{fluc}\phi_y$ (d) $\Lambda = 0°$ (e) $\Lambda = 10°$ (f) $\Lambda = 20°$ at $t/T = 0.12$. The color maps of the average and the fluctuating force density are different to highlight the difference in range in their plotting.**

Figure 8 shows an iso-surface plot of the normal force density field $-2\rho Q\phi_y$ for the three sweep angles at $t/T = 0.12$. For visualization purposes, the wings are rotated by the pitching angle. Figures 8 a, b & c show the phase averaged normal force distribution $-2\rho Q^{3D}_{avg}\phi_y$ for sweep angles $\Lambda = 0°$, $\Lambda = 10°$, & $\Lambda = 20°$ respectively. Similarly, Figures 8 (d-f) show the fluctuating Normal Force density field $-2\rho Q^{2D}_{fluc}\phi_y$ for those sweep angles. Owing to the distinct ranges of the mean and fluctuation components, each is visualized with a dedicated colormap optimized for its respective range. For the convenience of the reader, the cases for sweep angles $\Lambda = 0°$, $\Lambda = 10°$, & $\Lambda = 20°$ are boxed with blue, red, & green color, which is consistent with the color scheme throughout this manuscript. Although the normal force, pitching moment, and rolling moment all contribute to the CoP, only the normal force density is shown here for simplicity. The normal force is the key component in both span-wise and chord-wise movement of CoP, as revealed in equations 6 & 7. As the fluctuating term is calculated with the fluctuating component of the velocity field, it carries a signature of turbulence. At this instant ($t/T = 0.12$), the fluctuating term is not very significant compared to the phase-averaged term. Probably the coherent nature of the leading edge and wingtip vortex leads to less turbulence, which is reflected in the fluctuating terms.

The distributions of the forces and moments in the span-wise (*z*) and chord-wise (*x*) directions are obtained by integrating the vortical density term ($-2\rho Q\phi$) over the horizontal and vertical planes. For example, integration over the horizontal planes provides us with the force/moment distribution in the span-wise or *z* direction: $\int -2\rho Q(x,y,z)\phi(x,y,z)dxdy = F(z)$. Consequently, integration over the vertical plane results in chord-wise (x direction) distribution: $\int -2\rho Q(x,y,z)\phi(x,y,z)dydz = F(x)$. To explain the movement of CoP in the *z*-direction, we are interested in the span-wise distribution of the normal force $F_y(z)$ and rolling moment $M_x(z)$. Similarly, the chord-wise distribution of the normal force $F_y(x)$ and pitching moment $M_z(x)$ can be helpful in explaining the CoP movement in the *x*-direction. All the subsequent figures show the normalized distribution of the vortical terms either in the *x* or *z* direction. These distributions are normalized in the following way,

$$C_{F_y}(z) = \frac{F_y(z)}{\frac{1}{2}\rho U^2 c} \qquad C_{M_x}(z) = \frac{M_x(z)}{\frac{1}{2}\rho U^2 c^2} \qquad (30)$$

$$C_{F_y}(x) = \frac{F_y(x)}{\frac{1}{2}\rho U^2 s} \qquad C_{M_z}(x) = \frac{M_z(x)}{\frac{1}{2}\rho U^2 cs} \qquad (31)$$

The span-wise (*z*) distribution for the vortical terms of the normal force coefficient $C_{F_y}(z)$ and the rolling moment $C_{M_x}(z)$ are presented in the first row of the Figure 9 (a-c) and second row (d-f) at $t/T = 0.12$, respectively. The total (solid line), average (dashed line), and the fluctuating terms (dotted line) for the normal force are shown in Figure 9 a, b & c, respectively. The similar distributions for the rolling moment are shown in Figure 9 d, e & f. Both fluctuating and average components contribute to the total vortical term. Furthermore, the movement of the CoP primarily depends on the distribution of the total vortical term. The blue, red, & green colors represent the sweep angles $\Lambda = 0°$, $\Lambda = 10°$ & $\Lambda = 20°$, respectively. The span-wise distributions of the normal force and rolling moment are almost similar across

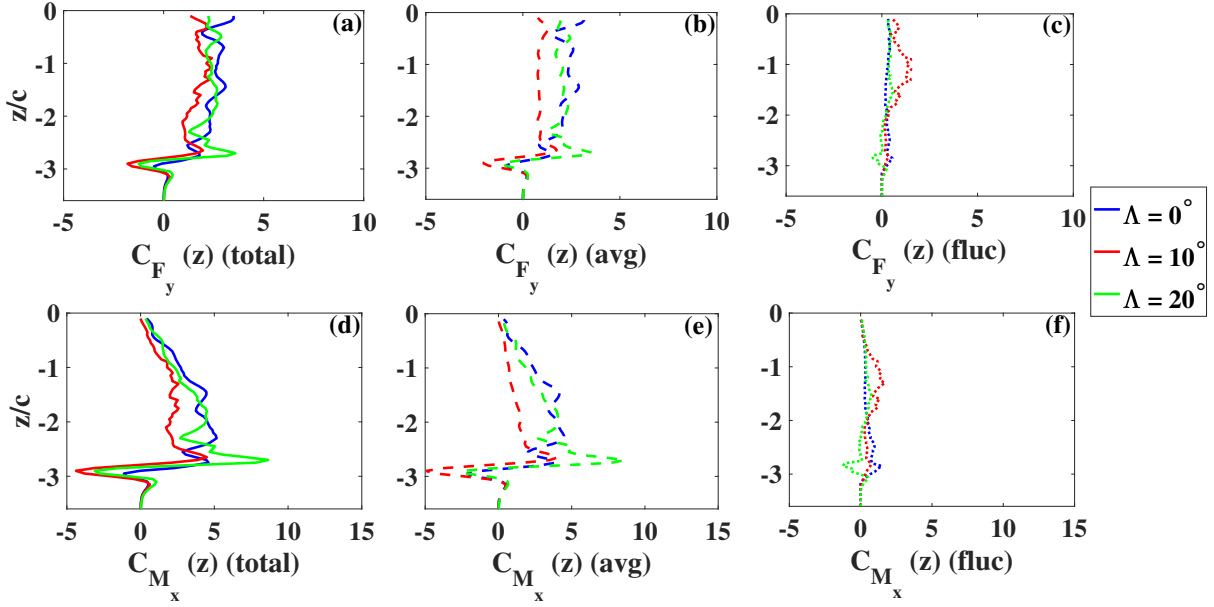

**Fig. 9** Coefficients of span-wise distribution of (a) total vortical term (b) average term (c) fluctuating term of the normal force, and (d) total vortical term (e) average term (f) fluctuating term of the rolling moment at $t/T = 0.12$

the sweep angles. Hence, the spanwise location of the center of pressure (CoP) does not vary significantly across the sweep angles as seen in Figure 7(d-f). It can be observed that the fluctuating terms (shown in Figure 9 c & f) are not as significant as the average vortical terms (shown in Figure 9 b & e), probably due to less turbulence shown in Figure 8.

The chord-wise ($x$) distribution for the vortical terms of the normal force coefficient $C_{F_y}(x)$ and the pitching moment $C_{M_z}(x)$ are presented in the first row of Figure 10 (a, b & c) and second row (d, e & f) at $t/T = 0.12$. The total, average, and fluctuating terms for the normal force are shown in Figures 10 a, b & c, respectively, and the similar distributions for the pitching moment are shown in Figures 10 d, e & f. The chord-wise distributions for the total vortical term of normal force and pitching moment contribute primarily to the position of the center of pressure in the chord-wise direction. The fluctuating terms for both normal force and the pitching moment, observed in Figure 10(e & f), are not significant compared to the average term shown in Figure 10(b & e). Hence, the primary contributor to the location of CoP in the chord-wise direction at $t/T = 0.12$ is the average term. For sweep angle $\Lambda = 0°$ & $\Lambda = 10°$, the significant variations of both the normal force (Figure 10a) and the pitching moment (Figure 10d) are more or less within the chord-length ($-0.5 < x/c < 0.5$), which keeps the CoP close to the mid-chord. However, for sweep angle $\Lambda = 20°$, there is relatively more variation of normal force and pitching moment beyond the trailing edge ($x/c > 0.5$), which pulls the CoP relatively further away from the mid-chord towards the trailing edge, as shown in Figure 7. This variation away from the mid-chord might primarily be due to the contribution from the wingtip vortex as observed in Figure

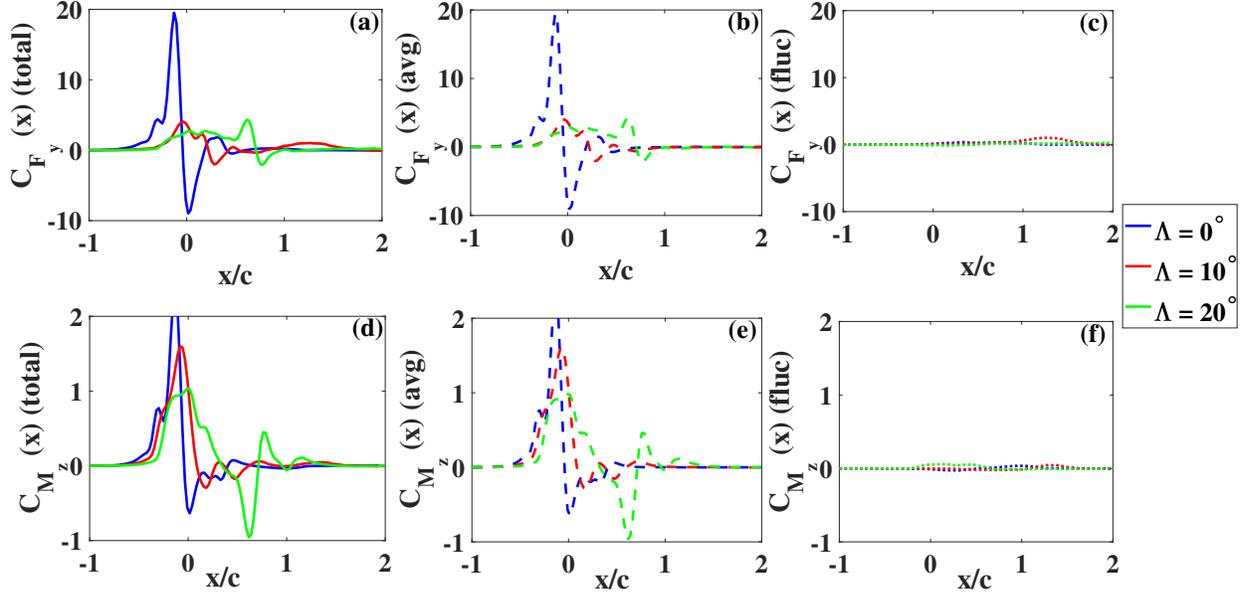

**Fig. 10** Coefficients of chord-wise distribution of (a) total vortical term (b) average term (c) fluctuating term of the normal force, and (d) total vortical term (e) average term (f) fluctuating term of the pitching moment at $t/T = 0.12$

8(c).

### E. Connection between Flow Structures and CoP at $t/T = 0.38$

Turning our attention to the time later in the cycle, Figure 11 shows an iso-surface plot of the normal force density field $-2\rho Q \phi_y$ for various sweep angles at $t/T = 0.38$. The figure locations with the alphabetical numbering and the color scheme are consistent with Figure 8. The average flow field for the sweep angle $\Lambda = 0°$ & $\Lambda = 10°$ shows the breakdown of the vortex structures in Figure 11(a) & (b), which was observed at time instant $t/T = 0.12$. The leading edge and wingtip vortex are also separated from the wing. However, there is still a vortex structure attached to the wing for sweep angle $\Lambda = 20°$ shown in Figure 11(c). The fluctuating flow structures in Figure 11(d-f) indicate an increase in turbulence signatures from the previous time instant $t/T = 0.12$, probably due to vortex breakdown.

The span-wise ($z$) distribution for the vortical terms of the normal force coefficient $C_{F_y}(z)$ and the rolling moment $C_{M_x}(z)$ are presented in the first row of Figure 12 (a, b & c) and second row of Figure 12 (d, e & f) at $t/T = 0.38$. The presentation structure is the same as Figure 9. The noteworthy difference of this time instant ($t/T = 0.38$) than the previous is the significant rise in the fluctuating term. For sweep angle $\Lambda = 0°$, there is a strong negative contribution to the normal force near the root ($z_{root}/c = 0$) of the wing. However, at the same location for sweep angle $\Lambda = 20°$, the total vortical term is strongly positive. These near-root contributions are primarily coming from the fluctuating terms as observed in Figure 12(c). However, these effects near the root are not seen for the rolling moment distribution in Figure

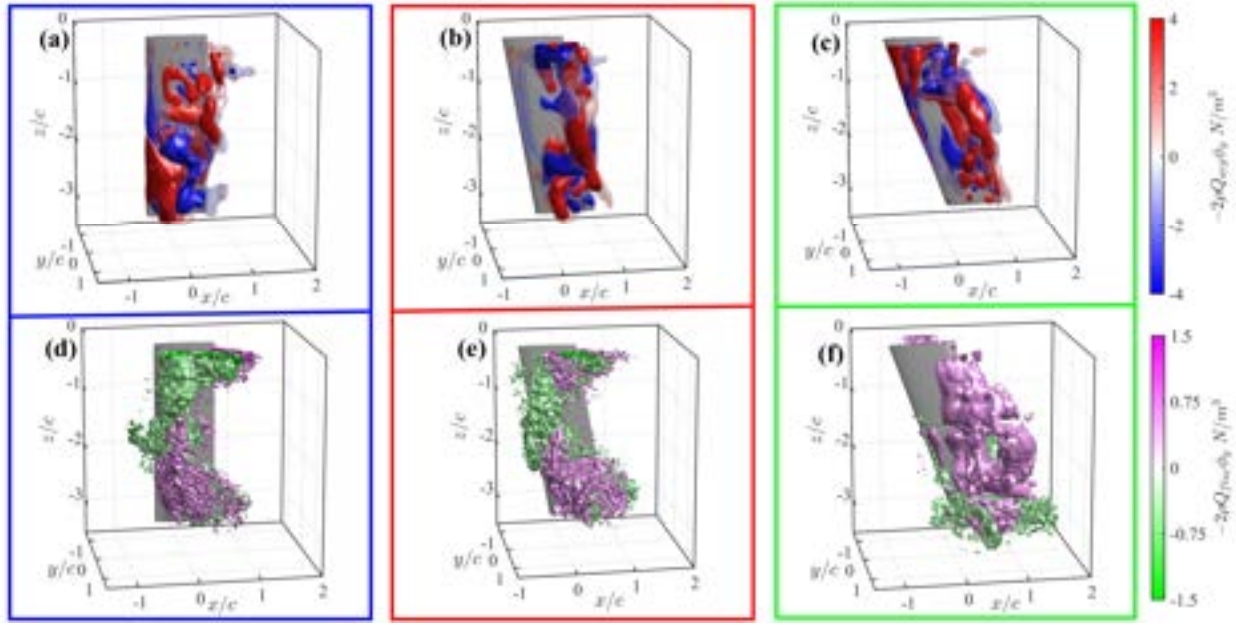

**Fig. 11 Iso surface plot of the average normal force density $-2\rho Q_{avg}\phi_y$ sweep angle (a) $\Lambda = 0°$ (b) $\Lambda = 10°$ (c) $\Lambda = 20°$ and the fluctuating normal force density $-2\rho Q_{fluc}\phi_y$ (d) $\Lambda = 0°$ (e) $\Lambda = 10°$ (f) $\Lambda = 20°$ at $t/T = 0.38$. The color maps of the average and the fluctuating force density are different to highlight the difference in range in their plotting, similar to Figure 8.**

12 (d-f) due to a smaller moment arm near the root. However, in both normal force and rolling moment, there is a strong positive contribution near the tip of the wing ($z_{tip}/c = -3$). Nonetheless, the primary factor that is different in this case for sweep angle $\Lambda = 0°$ & $\Lambda = 20°$ is the strong negative and positive contribution of the normal force near the root ($z_{root}/c = 0$). This difference explains the movement of the CoP for the wings with sweep angle $\Lambda = 0°$ & $\Lambda = 20°$ towards the wingtip and the wing root, respectively, as observed in Figure 7(d & f). The spanwise distribution of the normal force and the rolling moment of the sweep angle $\Lambda = 10°$ are similar to each other. Hence, the movement of the CoP for sweep angle $\Lambda = 10°$ is not significant in the span-wise direction as observed in Figure 7(e).

The chord-wise ($x$) distribution for the vortical terms of the normal force coefficient $C_{F_y}(x)$ and the pitching moment $C_{M_z}(x)$ are presented in the first row of Figure 13 (a, b & c) and second row of the Figure 13 (d, e & f) at $t/T = 0.38$. The presentation structure is the same as Figure 10. Similar to the span-wise distribution shown in Figure 12, the distribution of the fluctuating term is more significant than it was at time instant $t/T = 0.12$. For all the sweep angles, the fluctuating terms contribute to a strong positive pitching moment near $x/c = 1$, which pulls the CoPs towards the Trailing Edge (TE) as shown in Figure 7(a-c). However, this movement towards the TE is not the same across all sweep angles. For sweep angle $\Lambda = 20°$, the fluctuating and the average term of the pitching moment shown in Figure 13(e & f), are competing. And that translates to a smaller movement of the CoP for $\Lambda = 20°$ compared to the other sweep angles at this time instant.

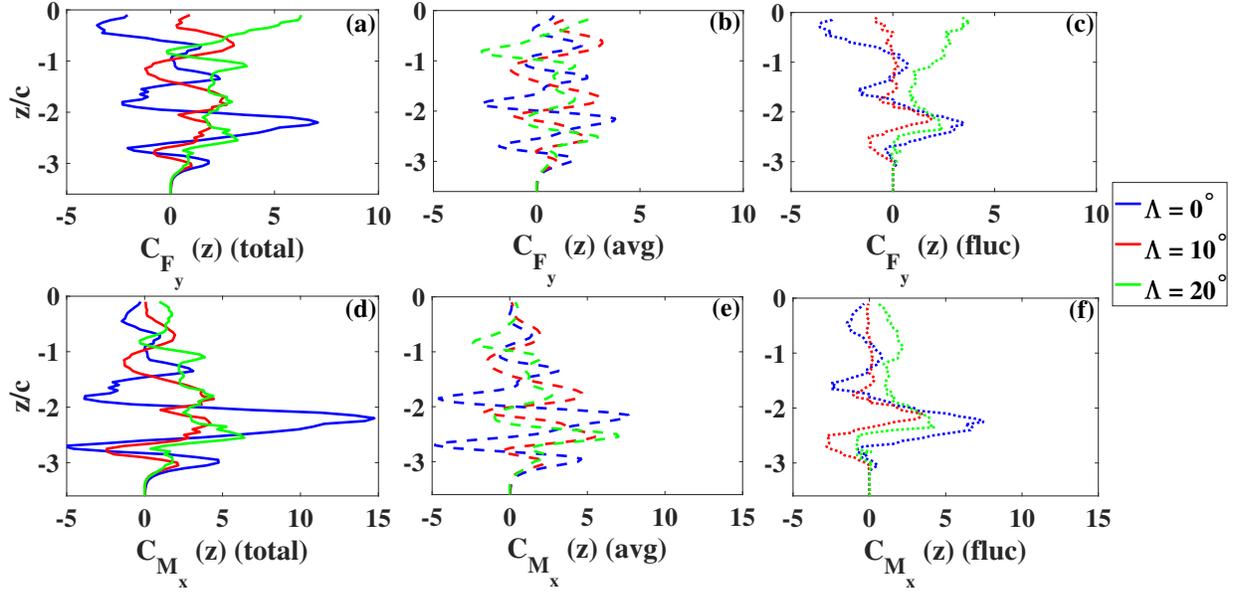

Fig. 12 Coefficients of span-wise distribution of (a) total vortical term (b) average term (c) fluctuating term of the normal force and (d) total vortical term (e) average term (f) fluctuating term of the rolling moment at $t/T = 0.38$

## V. Conclusion

The present work analyzes the center of pressure movement of rigid pitching swept wings. The wings underwent sinusoidal pitching motion with amplitude $A = 60°$ and frequency $f_p = 0.255 Hz$ (reduced frequency, $\kappa = 0.27$). Three wings with sweep angles $\Lambda = 0°$, $\Lambda = 10°$ & $\Lambda = 20°$ are considered for this study. The lift and drag measurement hardly shows any dependency on the sweep angle, but the pitching moment does. This observation motivated the calculation of the center of pressure (CoP) for all the cases. The calculation revealed that the movement of CoP in the cross-chord-wise direction is negligible compared to the chord-wise and span-wise directions. The CoP moves towards the wingtip and the trailing edge in case of $\Lambda = 0°$ & $\Lambda = 10°$. However, for sweep angle $\Lambda = 20°$, the CoP remains more or less at the same chord-wise location but moves towards the root of the wing in the spanwise direction.

We utilized the Force Moment Partitioning Method (FMPM) on the PIV data to explain the movement of CoP for various sweep angles. Since this is a vortex-dominated flow, all the terms (added mass, viscous stresses, & boundary terms) other than vortical terms are calculated to be negligible, and the total vortical terms generally show a good agreement with the measured forces and moments. The vortical term can also be resolved into two components: average and fluctuating terms. Two instances during the flapping cycle, $t/T = 0.12$ & $t/T = 0.38$, are chosen for further analysis to understand the movement of CoP. The force and moment density fields are integrated over the horizontal and vertical planes to obtain the distributions in the chord and spanwise directions. The strength of the fluctuating term increases at time $t/T = 0.38$ than it was at $t/T = 0.12$, probably because of the increase in turbulence due to vortex breakdown. The wingtip vortex has a significant role to play in normal force and rolling moment. The fluctuating flow field contributes to the spanwise movement of the CoP for $\Lambda = 0°$ & $\Lambda = 20°$. The fluctuating term is also responsible for pulling the CoP towards the trailing edge for sweep angle $\Lambda = 10°$. Accurate measurement of the CoP and analysis of the primary contributor to its movement can be helpful for the stability analysis. Moreover, capturing the locus of the CoP in a reduced-order model can help in the design of such efficient vehicles or structures.

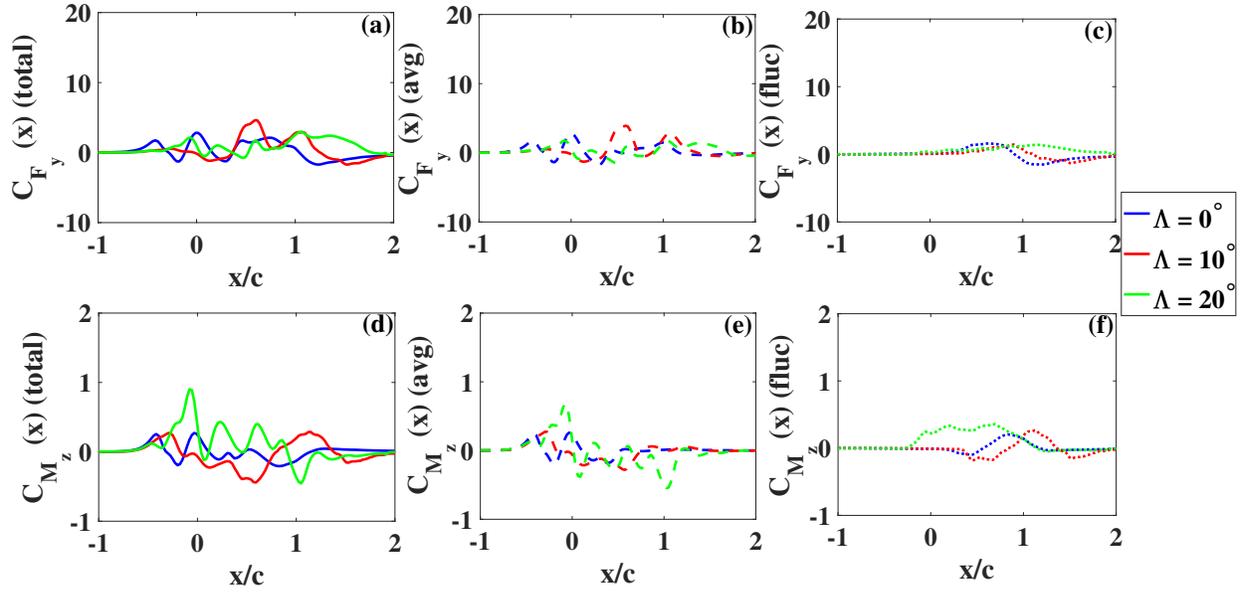

**Fig. 13** Coefficients of chord-wise distribution of (a) total vortical term (b) average term (c) fluctuating term of the normal force, and (d) total vortical term (e) average term (f) fluctuating term of the pitching moment at $t/T = 0.38$

In the end, the present study has some limitations that need to be highlighted. The CoP movement calculation in the region $t/T > 0.45$ has low confidence since both the forces and moments have very low values.

## Acknowledgments

This work was supported by the Air Force Office of Scientific Research, grant number FA9550-21-1-0462, monitored by Dr Gregg Abate.

## References

[1] Zhu, Y., and Breuer, K., "Flow-induced oscillations of pitching swept wings: stability boundary, vortex dynamics and force partitioning," *Journal of Fluid Mechanics*, Vol. 977, 2023, p. A1. https://doi.org/10.1017/jfm.2023.925.

[2] Rosenstein, H., and Clark, R., "Aerodynamic development of the V-22 tilt rotor," *Aircraft Systems, Design and Technology Meeting*, 1986, p. 2678.

[3] Larwood, S., Van Dam, C., and Schow, D., "Design studies of swept wind turbine blades," *Renewable Energy*, Vol. 71, 2014, pp. 563–571.

[4] Ellington, C. P., Van Den Berg, C., Willmott, A. P., and Thomas, A. L., "Leading-edge vortices in insect flight," *Nature*, Vol. 384, No. 6610, 1996, pp. 626–630.

[5] EJ, L. D. M. U. S., LLM, K. R. G. W. V., and JJ, H. P. H. A. V., "How swifts control their glide performance with morphing wings," *Nature Letters*, Vol. 446, 2007, pp. 1082–1085.

[6] Zurman-Nasution, A. N., Ganapathisubramani, B., and Weymouth, G. D., "Fin sweep angle does not determine flapping propulsive performance," *Journal of the Royal Society Interface*, Vol. 18, No. 178, 2021, p. 20210174.

[7] Borazjani, I., and Daghooghi, M., "The fish tail motion forms an attached leading edge vortex," *Proceedings of the Royal Society B: Biological Sciences*, Vol. 280, No. 1756, 2013, p. 20122071.

[8] Bottom Ii, R., Borazjani, I., Blevins, E., and Lauder, G., "Hydrodynamics of swimming in stingrays: numerical simulations and the role of the leading-edge vortex," *Journal of Fluid Mechanics*, Vol. 788, 2016, pp. 407–443.

[9] Polhamus, E. C., "Predictions of vortex-lift characteristics by a leading-edge suctionanalogy," *Journal of aircraft*, Vol. 8, No. 4, 1971, pp. 193–199.

[10] Chiereghin, N., Bull, S., Cleaver, D., and Gursul, I., "Three-dimensionality of leading-edge vortices on high aspect ratio plunging wings," *Physical Review Fluids*, Vol. 5, No. 6, 2020, p. 064701.

[11] Beem, H. R., Rival, D. E., and Triantafyllou, M. S., "On the stabilization of leading-edge vortices with spanwise flow," *Experiments in fluids*, Vol. 52, 2012, pp. 511–517.

[12] Wong, J. G., Kriegseis, J., and Rival, D. E., "An investigation into vortex growth and stabilization for two-dimensional plunging and flapping plates with varying sweep," *Journal of Fluids and Structures*, Vol. 43, 2013, pp. 231–243.

[13] Wong, J. G., and Rival, D. E., "Determining the relative stability of leading-edge vortices on nominally two-dimensional flapping profiles," *Journal of Fluid Mechanics*, Vol. 766, 2015, pp. 611–625.

[14] Onoue, K., and Breuer, K. S., "A scaling for vortex formation on swept and unswept pitching wings," *Journal of Fluid Mechanics*, Vol. 832, 2017, pp. 697–720.

[15] Wojcik, C. J., and Buchholz, J. H., "Vorticity transport in the leading-edge vortex on a rotating blade," *Journal of Fluid Mechanics*, Vol. 743, 2014, pp. 249–261.

[16] Visbal, M. R., and Garmann, D. J., "Effect of sweep on dynamic stall of a pitching finite-aspect-ratio wing," *AIAA Journal*, Vol. 57, No. 8, 2019, pp. 3274–3289.

[17] Taira, K., and Colonius, T., "Three-dimensional flows around low-aspect-ratio flat-plate wings at low Reynolds numbers," *Journal of Fluid Mechanics*, Vol. 623, 2009, pp. 187–207.

[18] Kim, D., and Gharib, M., "Experimental study of three-dimensional vortex structures in translating and rotating plates," *Experiments in Fluids*, Vol. 49, 2010, pp. 329–339.

[19] Hartloper, C., Kinzel, M., and Rival, D. E., "On the competition between leading-edge and tip-vortex growth for a pitching plate," *Experiments in fluids*, Vol. 54, 2013, pp. 1–11.

[20] Birch, J. M., and Dickinson, M. H., "Spanwise flow and the attachment of the leading-edge vortex on insect wings," *Nature*, Vol. 412, No. 6848, 2001, pp. 729–733.

[21] Zhang, K., Hayostek, S., Amitay, M., Burtsev, A., Theofilis, V., and Taira, K., "Laminar separated flows over finite-aspect-ratio swept wings," *Journal of Fluid Mechanics*, Vol. 905, 2020, p. R1.

[22] Zhang, K., Hayostek, S., Amitay, M., He, W., Theofilis, V., and Taira, K., "On the formation of three-dimensional separated flows over wings under tip effects," *Journal of Fluid Mechanics*, Vol. 895, 2020, p. A9.

[23] Ribeiro, J. H. M., Yeh, C.-A., Zhang, K., and Taira, K., "Wing sweep effects on laminar separated flows," *Journal of Fluid Mechanics*, Vol. 950, 2022, p. A23.

[24] Jasim, N. A., and Shamkhi, M. S., "The design of the center of pressure apparatus with demonstration," *Cogent Engineering*, Vol. 7, No. 1, 2020, p. 1843225.

[25] Chorin, A. J., "Numerical solution of the Navier-Stokes equations," *Mathematics of computation*, Vol. 22, No. 104, 1968, pp. 745–762.

[26] Adrian, R. J., Christensen, K. T., and Liu, Z.-C., "Analysis and interpretation of instantaneous turbulent velocity fields," *Experiments in fluids*, Vol. 29, No. 3, 2000, pp. 275–290.

[27] De Kat, R., and Van Oudheusden, B., "Instantaneous planar pressure determination from PIV in turbulent flow," *Experiments in fluids*, Vol. 52, 2012, pp. 1089–1106.

[28] Charonko, J. J., King, C. V., Smith, B. L., and Vlachos, P. P., "Assessment of pressure field calculations from particle image velocimetry measurements," *Measurement Science and technology*, Vol. 21, No. 10, 2010, p. 105401.

[29] Dabiri, J. O., Bose, S., Gemmell, B. J., Colin, S. P., and Costello, J. H., "An algorithm to estimate unsteady and quasi-steady pressure fields from velocity field measurements," *Journal of Experimental Biology*, Vol. 217, No. 3, 2014, pp. 331–336.

[30] Chen, J., Raiola, M., and Discetti, S., "Pressure from data-driven estimation of velocity fields using snapshot PIV and fast probes," *Experimental Thermal and Fluid Science*, Vol. 136, 2022, p. 110647.

[31] Quartapelle, L., and Napolitano, M., "Force and moment in incompressible flows," *AIAA journal*, Vol. 21, No. 6, 1983, pp. 911–913.

[32] Zhang, C., Hedrick, T. L., and Mittal, R., "Centripetal acceleration reaction: an effective and robust mechanism for flapping flight in insects," *PloS one*, Vol. 10, No. 8, 2015, p. e0132093.

[33] Moriche, M., Flores, O., and Garcia-Villalba, M., "On the aerodynamic forces on heaving and pitching airfoils at low Reynolds number," *Journal of Fluid Mechanics*, Vol. 828, 2017, pp. 395–423.

[34] Menon, K., and Mittal, R., "On the initiation and sustenance of flow-induced vibration of cylinders: insights from force partitioning," *Journal of Fluid Mechanics*, Vol. 907, 2021, p. A37.

[35] Menon, K., and Mittal, R., "Quantitative analysis of the kinematics and induced aerodynamic loading of individual vortices in vortex-dominated flows: a computation and data-driven approach," *Journal of computational physics*, Vol. 443, 2021, p. 110515.

[36] Menon, K., and Mittal, R., "Significance of the strain-dominated region around a vortex on induced aerodynamic loads," *Journal of fluid mechanics*, Vol. 918, 2021, p. R3.

[37] Li, J., and Wu, Z.-N., "Vortex force map method for viscous flows of general airfoils," *Journal of Fluid Mechanics*, Vol. 836, 2018, pp. 145–166.

[38] Li, J., Zhao, X., and Graham, M., "Vortex force maps for three-dimensional unsteady flows with application to a delta wing," *Journal of Fluid Mechanics*, Vol. 900, 2020, p. A36.

[39] Menon, K., Kumar, S., and Mittal, R., "Contribution of spanwise and cross-span vortices to the lift generation of low-aspect-ratio wings: Insights from force partitioning," *Physical Review Fluids*, Vol. 7, No. 11, 2022, p. 114102.

[40] Zhang, K., and Taira, K., "Laminar vortex dynamics around forward-swept wings," *Physical Review Fluids*, Vol. 7, No. 2, 2022, p. 024704.

[41] Zhu, Y., Lee, H., Kumar, S., Menon, K., Mittal, R., and Breuer, K., "Force moment partitioning and scaling analysis of vortices shed by a 2D pitching wing in quiescent fluid," *Experiments in Fluids*, Vol. 64, No. 10, 2023, p. 158.

[42] Menon, K., and Mittal, R., "Quantitative analysis of the kinematics and induced aerodynamic loading of individual vortices in vortex-dominated flows: A computation and data-driven approach," *Journal of Computational Physics*, Vol. 443, 2021, p. 110515. https://doi.org/https://doi.org/10.1016/j.jcp.2021.110515, URL https://www.sciencedirect.com/science/article/pii/S0021999121004101.

# Appendices

## A. Influence Potential

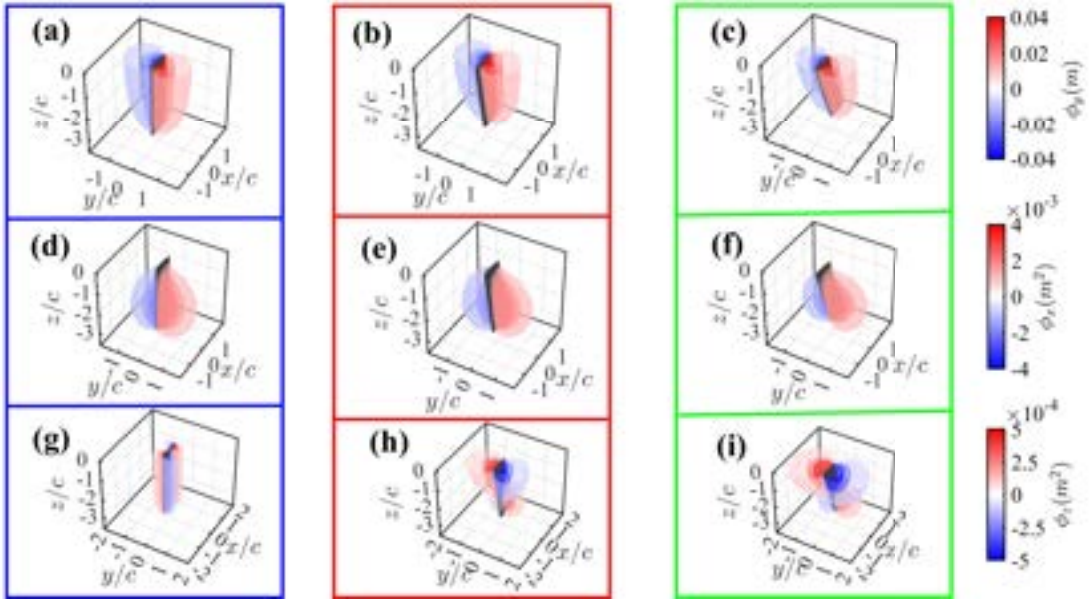

**Fig. 14** Influence potential for Normal Force (a-c), Rolling Moment (d-f), Pitching Moment (g-i)

The Figure 14 shows the influence potential of the normal force ($\phi_y$), rolling moment ($\phi_x$) and the pitching moment ($\phi_z$). The row shows the influence potential for a particular force/moment, and each column represents a sweep angle. The influence potential for sweep angles $\Lambda = 0°$, $\Lambda = 10°$, and $\Lambda = 20°$ are enveloped in blue, red, and green boundaries, respectively. The Figures 14 a, b & c show the Normal Force influence of sweep angles $\Lambda = 0°$, $\Lambda = 10°$, and $\Lambda = 20°$. The rolling and pitching moment influence potentials follow a similar pattern shown in Figures 14 (d-f) and Figure 14 (g-i), respectively.

## B. Q condition

In the present work, due to the limitations of planar stereo PIV, FMPM estimation of the forces and moments is calculated using the full three-dimensional $Q$ field of the coherent phase-averaged flow: $Q^{3D}_{avg}$, but only the two-dimensional $Q$ field associated with the fluctuating velcicty gradients, $Q^{2D}_{fluc}$. However, we obtained good agreement with measured forces and moments and the FMPM estimation, even with $Q^{2D}_{fluc}$. Here we justify this simplification. One property of the $Q$ field is that, for incompressible flow, $I = \int Q dV = 0$ at each instant in time [36]. In the present study, $Q = Q^{3D}_{avg} + Q^{2D}_{fluc}$ (equation 23), and Figure 15 shows the integrals, $I^{3D}_{avg} = \int Q^{3D}_{avg} dV$ (in purple), $I^{2D}_{fluc} = \int Q^{2D}_{fluc} dV$ (in yellow), & $I = I^{3D}_{avg} + I^{2D}_{fluc}$ (in black) for the half cycle. Figure 15 (a-c) shows the integrals for sweep angles 0°, 10°, and 20°, in blue, red, and green, respectively, consistent with the color scheme of this manuscript. The data is normalized by $J = (\int_0^T \int (|Q_{avg}^{3D}| + |Q_{fluc}^{2D}|) dV dt)/T$. The normalized data represent the integrals as a ratio of the time average of the total $Q (= |Q_{avg}^{3D}| + |Q_{fluc}^{2D}|)$, for the respective sweep angles. If all of the $Q$ field were fully accounted, this sum would equal zero. It can be noted that the total integral ($I = I_{avg}^{3D} + I_{fluc}^{2D}$) is less than 20% of the total time-averaged $Q$, for sweep angle $\Lambda = 10°$ & $\Lambda = 20°$. Only for the sweep angle $\Lambda = 0°$, this variation is less than 10%, which might explain the better agreement of FMPM with the measured forces and moments for $\Lambda = 0°$ than $\Lambda = 20°$ & $\Lambda = 10°$, shown in Figure 6.

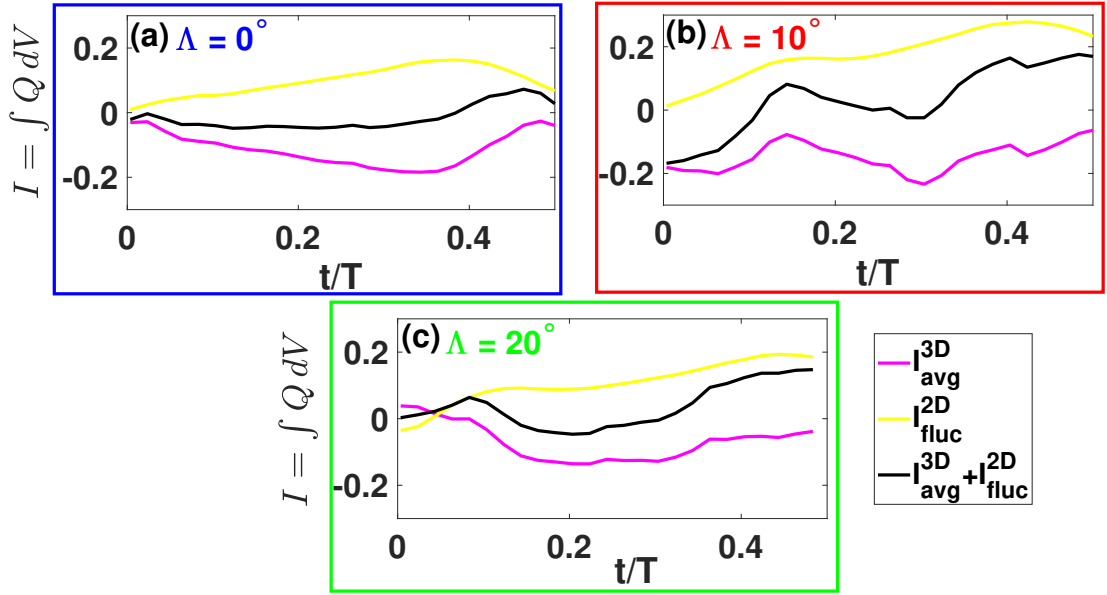

**Fig. 15** Integral of 3D phase-averaged $Q_{avg}^{3D}$ and 2D fluctuating $Q_{fluc}^{2D}$ for various sweep angles at a given time during the pitching cycle.